\documentclass[a4paper,11pt]{article}

\begin{document}
\title{Continuous Spin Representation from Contraction of the Conformal Algebra}

\author{Abu Mohammad Khan\footnote{Email: abu.khan@bracu.ac.bd, \ abu.m.khan@gmail.com}\\ Department of Computer Science \& Engineering \\ BRAC University\\ 46 Mohakhali, Dhaka -1212,  Bangladesh}

\date{February 08, 2021}

\maketitle
\begin{abstract}
In this paper, we discuss the In\"on\"u-Winger contraction of the conformal algebra. We start with the light-cone form of the Poincar\'e algebra and extend it to write down the conformal algebra in $d$ dimensions. To contract the conformal algebra, we choose five dimensions for simplicity and compactify the third transverse direction in to a circle of radius $R$ following Kaluza-Klein dimensional reduction method. We identify the inverse radius, $1/R$, as the contraction parameter. After the contraction, the resulting representation is found to be the continuous spin representation in four dimensions. Even though the scaling symmetry survives the contraction, but the special conformal translation vector changes and behaves like the four-momentum vector. We also discussed the generalization to $d$ dimensions.
\end{abstract}


\section{Introduction}\label{intro}
Recently the interest in continuous spin representation (CSR) of the Poincar\'e group has grown tremendously. The eigenvalues of the Casumir operators for these repesentations in $d$ dimensions are the momentum squared which is zero and the non-zero values of the squares of the Pauli-Lubansky forms\cite{Brink:2002zx}. In four dimensions, these are the four-momentum squared and the square of the Pauli-Luba\'nski vector. These representations are also known as the Wigner's infinite spin representation in the literature. The CSRs did not attract much attention until recently mainly because no known particles seem to obey this representation, and also because of the `No-Go Theorem' by Weinberg and Witten\cite{Weinberg:1980kq}.

However, in 2002 Brunetti {\it et.al.}\cite{Brunetti:2002nt} showed the possibility of localization of the CSR fields in spacelike cones. This observation renewed the interest in CSR of the Poincar\'e group. In 2005, Mourad\cite{Mourad:2005rt} showed that the CSRs may arise in the context of String theory. Metsaev\cite{Metsaev:2017ytk,Metsaev:2019opn,Metsaev:2016lhs,Khabarov:2017lth} argued their existence in different contexts. In 2015, Longo {\it et.al.}\cite{Longo:2015tra} and Schroer\cite{Schroer:2015rct} showed that the CSRs may have only non-point like localization as in String Theory. A beautiful discussion on incompatibility of CSRs with point-like localization is presented by K\"{o}hler\cite{Kohler:2015cua}. Rehren\cite{Rehren:2017xzn} also constructed a string-localized infinite spin quantum stress-energy tensor even without a classical action.

On the other hand, Schuster and Toro\cite{Schuster:2013pta,Schuster:2014hca} proposed the action principle for the particles with integer CSRs, and later Bekaert {\it et.al.}\cite{Najafizadeh:2015uxa} presented the action for particles with half-odd integer CSRs. Very recently Najafizadeh\cite{Najafizadeh:2019mun} presented the supersymmetry transformation to construct the supermultiplets associated with the CSRs. Buchbinder {\it et.al.}\cite{Buchbinder:2018yoo,Buchbinder:2019iwi} developed the BRST method to construct the Lagrangian for bosonic fields. They also obtained the field realization of the infinite spin $N=1$ supersymmetry. Using Kirillov method, Gracia-Bond\'ia {\it et.al.}\cite{Gracia-Bondia:2017fai} quantized the fields. These developments and other approaches has been concisely summarized by Bekaert and Skvortsov in a recent article\cite{Bekaert:2017khg}.

The rapid progress in the area of CSR after the work of Brunetti {\it et.al.} is very encouraging for us and it motivates us to think how do these infinite spin representations arise from a theory that has conformal symmetry or AdS symmetry, like String Theory. It is very well known for a long time that the Conformal group contains the Poincar\'e group as a sub group, and it only admits regular massless representations of the Poincar\'e group\cite{Angelopoulos:1980wg, Khan:2004nj, Yao:1971xg}. This clearly states that the CSR can not be obtained by any non-singular transformations. Hence, only other possibility is to consider a singular transformation, like group contraction, to find out if the conformal group give rise to the CSR in lower dimensions.

In an earlier paper\cite{Khan:2004nj}, we discussed how the CSR can be obtained in lower dimensions by contracting the Poincar\'e group in higher dimensions. In that paper, we used the In\"on\"u-Winger group contraction method\cite{Inonu:1953sp} by considering the inverse radius of the Kaluza-Klein(KK) compactification\cite{1921SPAW.......966K, 1926ZPhy...37..895K} as the contraction parameter. It is also very well-known that the contraction of AdS groups\cite{AyalaSanchez:2002wm} give rise to the Poincar\'e group with massless representation in lower dimensions. However, the contraction of the conformal group has not yet been discussed. Naively it is not surprising to expect that the contraction of the conformal group would only yield regular massless representations. However, in this paper, we have found that it gives rise to the CSR as well.

In this paper, we first consider the conformal algebra in five dimensions in light-cone form, apply the Kaluza-Klein(KK) dimensional reduction, followed by the IW contraction to get the generators in four dimensions. We show that after the contraction, the conformal symmetry is lost. In fact, the special conformal translation vectors become a linear combination of the four-momentum vectors and the  light-cone translation vectors. The resulting algebra and the symmetry properties are those of the continuous spin representations of the Poincar\'e group in four dimensions. We also discuss the generalization to $d$ dimensions.

We organize the paper as in the following ways: in Section-\ref{light-cone}, we review the light-cone formulation of the Poincar\'e algebra and its conformal extension, in Section-\ref{cont-confalg}, we develop the contraction parameter by KK compactification and use it to contract the conformal algebra to obtain the CSR, and finally in Section-\ref{disc}, we discuss the result and make the concluding remarks.
\section{The Light-cone form of the Conformal algebra}\label{light-cone}
In this section, we briefly present the conformal algebra in the light cone form. We begin with the Poincar\'e algebra in $d$ dimensions and then extend it by adding generators of scaling and special conformal transformations to complete the conformal algebra in $d$ dimensions. Since the conformal symmetry requires the particles to be massless, only the massless representation of the Poincar\'e group can be embedded in the conformal representation.
\subsection{The light-cone form of the Poincar\'e algebra}
In this subsection, we present a short review on the light-cone form of the Poincar\'e algebra introduced by Dirac\cite{RevModPhys.21.392} and its isomorphic form in a infinite momentum frame shown by Bacry and Chang\cite{osti_4502829}.

In $d$-dimensions, the Poincar\'e group generators, $P^\mu$ and $M^{\mu \nu}$, satisfy the following commutation relations,
\begin{eqnarray}
  \left[ \, P^\mu \, , \, P^\nu \, \right] &=& 0 \ , \label{eq1} \\
  \left[ \, M^{\mu \nu} \, , \, P^\alpha \, \right] &=& i \left( \eta^{\mu \alpha} P^\nu - \eta^{\nu\alpha} P^\mu \right) \ , \label{eq2} \\
  {\rm and} \quad \left[ \, M^{\mu\nu}\, , \, M^{\alpha\beta}\,\right] & = & i \left( \eta^{\mu\alpha} M^{\nu\beta} + \eta^{\alpha\nu} M^{\beta\mu} + \eta^{\nu\beta} M^{\mu\alpha}  + \eta^{\beta\mu} M^{\alpha\nu}\right) \ , \label{eq3}
\end{eqnarray}
where $\mu, \nu, \alpha, \beta = 0, 1, 2, \cdots , (d-1)$, and the metric signature is $\eta^{\mu \nu} = - ++\cdots +$. The representations are characterized by the eigenvalues of the Casimirs: $P^2\equiv P^\mu P_{\mu}$ and the squares of the Pauli-Luba\'nski $k$-forms which are given by\cite{Brink:2002zx}
$$ W_{(k)} ~ = ~ \frac{\epsilon_{\mu_1\mu_2\cdots \mu_k \mu_{k+1} \mu_{k+2} \mu_{k+2} \cdots \mu_{d-1} \mu_{d}} P^{\mu_{d}} M^{\mu_{k+1}\nu_{k+2}}\cdots M^{\mu_{d-2}\mu_{d-1}}}{\sqrt{k! 2^{(d-k+1)/2}((d-k-1)/2)((d-k-1)/2)!}} \ , $$
where $k=1,3,\cdots, (d-3)$ if $d$ is even and $k=0,2,\cdots, (d-3)$ if $d$ is odd. In momentum space the canonical representation of the rotation and boost generators of the Poincar\'e group are given by
\begin{eqnarray}
M^{ij} & = & -i \left( p^i \partial_ p^j - p^j \partial_ p^i \right) + S^{ij}  \ , \label{rot} \\ {\rm and} \ \ M^{0i} & = & -i p^0 \partial_p^i + \left( \frac{1}{-p^0 + m }\right) p^j S^{ij} \ , \label{boost}
\end{eqnarray}
respectively, where $i,j = 1, \cdots , (d-1)$ and
$$ \partial^i_p ~ = ~ \frac{\partial}{\partial p_i} \ . $$
The generators $S^{ij}$ satisfy the following commutation relation
$$ \left[ \,S^{ij} \, , \, S^{mn} \,\right] ~ = ~ i \left( \eta^{im} S^{jn} + \eta^{mj} S^{ni} + \eta^{jn} S^{im} + \eta^{ni} S^{mj}\right) \ . $$

For simplicity, we restrict ourselves to $d=4$ dimensions. Following Dirac, we introduce the light-cone coordinates
$$ x^\pm = \frac{1}{\sqrt{2}} \left( x^0 \pm x^{3} \right)  \quad {\rm and} \quad p^\pm = \frac{1}{\sqrt{2}} \left( p^0 \pm p^{3} \right)  \ . $$
The commutation relations now read
$$ \big[ \, x^- \, , \, p^+ \, \big] = -i \quad {\rm and} \quad \big[ \, x^a \, , \, p^b \, \big] = i \eta^{ab} \ , $$
where $a,b=1,2$ are the transverse dimensions. The momentum generators become,
\begin{equation} \label{mass-shell} P^+ = p^+, \quad P^a = p^a \quad {\rm and} \quad P^- = \frac{p^a p^a + m^2}{2p^+} \ . \end{equation}
Here $P^-$ is called the light-cone Hamiltonian. The six spacetime generators of the Poincar\'e group now split up as
$$ M^{\mu\nu} = \left( M^{ij} \, , \, M^{0i} \right) \to \left( M^{+-}, \ M^{ab}, \ M^{+a} , \ M^{-b}\right) \ , $$
where $i,j= 1,2,3$. Using the canonical representations in Eqs.(\ref{rot},\ref{boost}), we now compute the light-cone forms of the generators. The $M^{+-}$ generator becomes,
\begin{eqnarray} \nonumber
    M^{+-} & = & -M^{03} \ , \\ & = & i p^+ \partial_+ - \frac{\sqrt{2} p^a S^{3a}}{p^+ + p^- - \sqrt{2} \,m} \ ,\label{m_pm}
\end{eqnarray}
where we used
\begin{equation}\label{constraint} \frac{\partial}{\partial p^3} = \frac{\sqrt{2} p^+}{p^+ + p^-} \frac{\partial}{\partial p^+} \ , \end{equation}
because of the mass-shell condition in Eq.(\ref{mass-shell}).
The light-cone Hamiltonian, $M^{-a}$, becomes
\begin{eqnarray}
M^{-a} & = & \frac{1}{\sqrt{2}} \left( M^{0a} - M^{3a}\right) \ , \nonumber\\
& = &  \frac{1}{\sqrt{2}} \left[ -ip^0\partial^a_p + \left(\frac{1}{-p^0 + m}\right) \, p^j S^{aj} + i \left( p^3 \partial_p^a - p^a \partial_p^3 \right) - S^{3a} \right] \ ,  \nonumber \\
& = & - i  p^- \partial_p^a - \frac{i}{\sqrt{2}} \, p^a \partial_p^3 - \frac{1}{\sqrt{2}} \left( \frac{-p^0 + p^3 +m}{-p^0 + m}\right) S^{3a} + \frac{1}{\sqrt{2}} \,\frac{p^bS^{ab}}{-p^0 +m} \ , \nonumber \\
& = & -i p^- \partial^a_p  - i\frac{p^a p^+}{p^+ + p^-} \partial^+_p - \left( \frac{\sqrt{2}p^- -m}{p^+ + p^- - \sqrt{2} m}\right) S^{3a} - \frac{p^b S^{ab}}{p^+ + p^- - \sqrt{2} m} \ . \label{m-minusa}
\end{eqnarray}
The generators, $M^{+a}$, also change to the following forms,
\begin{equation} \label{m-plusa}
M^{+a} ~ = ~  -i p^+ \partial^a_p  + i\frac{p^a p^+}{p^+ + p^-} \partial^+_p + \left(\frac{\sqrt{2}p^+ -m}{p^+ + p^- - \sqrt{2} m}\right) S^{3a} - \frac{p^b S^{ab}}{p^+ + p^- - \sqrt{2} m} \ .
\end{equation}
The transverse rotation generators, $M^{ab}$, do not change. We now apply a boost along the third direction. The generators are mapped to the boosted frame and scaled in the following ways
\begin{eqnarray}
p^\pm \to \widehat{p}^{\,\,\pm} & = & e^{i\phi M^{03}} p^\pm e^{-i\phi M^{03}} = e^{\pm \phi} p^\pm \ , \\
p^a \to \widehat{p}^{\,\,a} & = & e^{i\phi M^{03}} p^a e^{-i\phi M^{03}} = p^a \ , \\
M^{\pm a}\to \widehat{M}^{\,\,\pm a} & = & e^{i\phi M^{03}} M^{\pm a} e^{-i\phi M^{03}} = e^{\pm \phi} M^{\pm a}\ , \\
M^{ab}\to \widehat{M}^{\, ab} & = & e^{i\phi M^{03}} M^{ab} e^{-i\phi M^{03}} = M^{ab}\ , \\
{\rm and} \ \ M^{+-}\to \widehat{M}^{\, +-} & = & e^{i\phi M^{03}} M^{+-} e^{-i\phi M^{03}} = M^{+-}\ , \end{eqnarray}
where the rapidity parameter $\phi$ is related to the speed $v$ of the boost by the following relation,
$$ e^\phi ~ = ~ \sqrt{\frac{1+v}{1-v}} \ . $$
To go to the infinite momentum frame, we take the limit $\phi \to \infty$. In this limit, the canonical representations of the Poincar\'e algebra in light-come form in Eqs.(\ref{mass-shell},\ref{m_pm},\ref{m-minusa},\ref{m-plusa}) and the transverse generator reduce to the following simple forms respectively (expressed in the usual position space),
\begin{eqnarray}
\widehat{P}^{\,+} & = & \widehat{p}^{\,+} , \quad \widehat{P}^{\,a} = \widehat{p}^{\,a} , \quad \widehat{P}^{\,-} = \frac{(\widehat{p}^{\,a})^2 + m^2 }{2\widehat{p}^{\,}+} \ , \label{hamiltonian}\\
\widehat{M}^{\,\,+-} &=& -\widehat{x}^{\,\,-} \, \widehat{p}^{\,\,+} \ ,  \label{boost2}\\
\widehat{M}^{\,\,-a} & = &  \widehat{x}^{\,\,-} \widehat{p}^{\,\,a} -\frac{1}{2} \{ \widehat{x}^{\,\,a} \, ,\, \widehat{P}^{\,\,-}\} + \frac{1}{\widehat{p}^{\,+}} ( \widehat{T}^a - \widehat{p}^{\,b} S^{ab}) \ , \label{boost4} \\
\widehat{M}^{\,\,+a} &=& -\widehat{x}^{\,a}\,\, \widehat{p}^{\,+} \ ,  \label{boost1}\\
\widehat{M}^{\,\,ab} &=& \widehat{x}^{\,a}\,\, \widehat{p}^{\,b} - \widehat{x}^{\,b}\,\, \widehat{p}^{\,a} + S^{ab}\ . \label{boost3}
\end{eqnarray}
Here the light-cone translation vector is defined as $\widehat{T}^a \equiv m S^{3a}$. The generalization to $d$ dimensions can be obtained by extending the transverse direction upto $(d-2)$, {\it i.e.} $a,b= 1, 2, \cdots, (d-2)$. The transverse generators $S^{ab}$ and the light-cone vectors $\widehat{T}^{\,a}$ form the transverse little group $SO(d-2)$ and satisfy the following algebra
\begin{eqnarray*}
  \left[ \, S^{ab} \, , \, S^{cd} \, \right]  &=& i\left( \eta^{ac} S^{bd} + \eta^{cb} S^{da} + \eta^{bd} S^{ac} + \eta^{da} S^{cb} \right)\ , \\
  \left[ \, S^{ab} \, , \, \widehat{T}^c \,\right]  &=& i \left( \eta^{ac} \widehat{T}^b - \eta^{bc} \widehat{T}^a \right) \ ,  \\
  {\rm and} \ \ \left[ \, \widehat{T}^a \, , \, \widehat{T}^b \,\right]  &=& im^2 S^{ab} \ .
\end{eqnarray*}
For massless particles, we put $m=0$ and $\widehat{T}^a$'s become commuting vectors. The states can be labelled by two ways:
\begin{itemize}
\item $\widehat{T}^a=0$. This is the regular massless representation. All known massless particles belong to this representation. The states satisfy $\widehat{T}^a |\psi > = 0$, and are labelled by the $SO(d-2)$ little group, in addition to $\widehat{p}^{\,+}$ and $\widehat{p}^{\,i}$.
\item $\widehat{T}^a\ne 0$. This is the CSR. The states satisfy $\widehat{T}^a|\psi> = \xi^a|\psi>$, and are labelled by the short little group $SO(d-3)$ that leaves $\widehat{T}^a$ invariant, $\widehat{p}^{\,+}$, $\widehat{p}^{\,a}$ and the length of the light-cone vector, $\xi^a$.
\end{itemize}
In the following, we add the scaling operator and the special conformal translation operator to the Poincar\'e algebra, and complete the conformal algebra relations in $d$ dimensions in light-cone forms. We also drop the `{\it hat}' symbols to represent the generators in the infinite momentum frame for simplicity.
\subsection{Conformal extension of the Poincar\'e algebra in light-cone form}\label{conf-ext}
The conformal symmetry is the symmetry for massless particles. Hence it is very natural to assume that the massless representations of the Poincar\'e group also have conformal symmetry. Since the Conformal group is larger than the Poincar\'e group, the massless representations of the Poincar\'e group can reside inside the conformal group as an embedding.

The scaling operator $D$, also known as the Dilatation operator, and the special conformal translation vector $K^\mu$ are given by
\begin{eqnarray*}
   D &=& \frac{1}{2} \big( x\cdot p + p \cdot x\big), \\
 {\rm and} \quad K^\mu  &=& 2x^\nu M^{\mu \nu} + x^2 p^\mu = 2 x^\mu D + 2x_\nu S^{\mu\nu} - x^2 p^\mu \ ,
 \end{eqnarray*}
 where $\mu,\nu=0,1,2, \cdots , (d-1)$, are added to the generators of the Poincar\'e group. In $d$ dimensions, these two generators satisfy the following commutation relations\cite{DiFrancesco:1997nk},
 \begin{eqnarray}
  \left[ \,  M^{\mu \nu} \, , \, D \, \right]   &=& 0 \ ,  \label{eq4} \\
  \left[ \, D\, , \, p^\mu \, \right]  &=& i p^\mu \ , \label{eq5} \\
  \left[ \, D\, , \, K^\mu \, \right]  &=& -i K^\mu \ , \label{eq6} \\
 \left [ \, p^\mu \, , \, K^\nu \, \right]  &=&  -2i \big( \eta^{\mu \nu} - M^{\mu \nu} \big)  \ , \label{eq7}\\
 {\rm and} \quad  \left[ \, M^{\mu \nu} \, , \, K^\alpha \, \right]  &=& i \big( \eta^{\mu\alpha} K^\nu - \eta^{\nu \alpha} K^\mu \big) \ . \label{eq8}
 \end{eqnarray}
 The commutation relations in Eqs.(\ref{eq1}-\ref{eq3}) and together with the commutation relations in Eqs.(\ref{eq4}-\ref{eq8}) complete the algebra of the Conformal group in $d$ dimensions.

 To obtain the light-cone form, we first express the generators, $D$ and $K^\mu$, in light-cone coordinates, and apply a boost by setting $x^+=0$. We use the expressions from Eqs.(\ref{mass-shell}-\ref{boost4}) to write down the light-cone forms of these generators and get,
 \begin{eqnarray}
 D & = & \frac{1}{2} \left( - \left\{ x^- , p^+ \right\}  + \left\{ x^a , p^a \right\} \right) \ , \label{dilatation} \\
 K^+ & = & - x^a x^a  p^+ \ , \label{k-plus} \\
 K^- & = & 2x^- D - x^a x^a P^-  + \frac{2}{p^+} \left( x^a T^a - x^a p^b S^{ab} \right) \ , \label{k-minus} \\
 {\rm and} \quad K^a & = & 2x^a D - x^c x^c p^a + 2x^b S^{ab} \ ,\label{k-transverse}
 \end{eqnarray}
 where $a,b = 1,2$. Now we compute the following commutators,
 \begin{eqnarray*}
   \big[ \, p^a \, , \, K^- \, \big]  &=& -2i \left( x^- p^a - \frac{1}{2}\left\{ \, x^a\, , \, P^- \,\right\} + \frac{1}{p^+} \left( T^a - p^b S^{ab} \right) \right) \ , \\
  {\rm and} \quad \big[ \, P^- \, , \, K^a\, \big]   &=& 2i \left( x^- p^a - \frac{1}{2}\left\{ \, x^a\, , \, P^- \,\right\} - \frac{1}{p^+} \left(  p^b S^{ab} \right) \right) \ .
 \end{eqnarray*}
 To satisfy the commutation relation in Eq.(\ref{eq7}), the right-hand sides of the above two equations must be negative of each other. This is possible only if $T^a=0$ in Eq.(\ref{boost4}), and hence in Eq.(\ref{k-minus}). This clearly indicates that only the regular massless representation of the Poincar\'e algebra has conformal extension. The light-cone boost now becomes,
 $$ M^{-a} ~ = ~  x^- p^a -\frac{1}{2} \{ x^a \, , \, P^-\} -\frac{1}{p^+} \,  p^b S^{ab} \ . $$
The above results are, of course, not new. We present here for clarity and completeness of our work in this paper.

In the following section-\ref{cont-confalg}, we will contract the conformal representation. Even though only the regular massless representation has conformal extension, the contraction may provide different massless representation, namely the CSR.
\section{Contraction of the Conformal algebra}\label{cont-confalg}
In this section, we contract the conformal algebra following the procedure introduced by In\"on\"u and Wigner\cite{Inonu:1953sp}.  Here we proceed by compactifying a direction on to a circle of radius $R$, known as the KK dimensional reduction method\cite{1921SPAW.......966K, 1926ZPhy...37..895K}. The inverse radius, $1/R$, is considered to be the contraction parameter. When the contraction limit, $R\to \infty$ or equivalently $1/R \to 0$ is applied, we obtain the CSR. In the following, we first very briefly review the IW contraction method as described in their paper\cite{Inonu:1953sp}, and then apply the KK reduction to a transverse direction to contract the algebra.
\subsection{In\"on\"u-Winger contraction: A brief review}
In their paper\cite{Inonu:1953sp}, In\"on\"u and Winger obtained the inhomogeneous Euclidean group $E_2$ by contracting the homogeneous $SO(3)$ group. The rotation generators of the $SO(3)$ group are $L_{ij}$ for $i,j=1,2,3$. Defining $L_i = \frac{1}{2}\epsilon_{ijk} L_{jk}$, the generators satisfy the relation,
\begin{eqnarray*}
\big[ \, L_i \, , \,L_j\, \big] & = & i\epsilon_{ijk} L_k \, \\
  L^2_{1} + L^2_{2} + L^2_{3}   &=& l(l+1) \ ,  \\
  L_{3}  &=& m \ ,
\end{eqnarray*}
when acts on a state. The representation space is $l(l+1)$ dimensional. Now we define, $\widehat{L}_i \equiv \epsilon L_i$, where $\epsilon$ is an arbitrary contraction parameter. For a fixed $m$, we have,
$$ \widehat{L}^2_1 + \widehat{L}^2_2 + \widehat{L}^2_3 ~ = ~ \epsilon^2 l(l+1) \ . $$
Now if we take the limits, $\epsilon\to 0$ and $l\to \infty$ such that $\epsilon l\equiv \Xi = \mbox{fixed}$, we get,
$$ \widehat{L}^2_1 + \widehat{L}^2_2 ~ = ~ \Xi^2 \ , $$
where we used the fact that $\widehat{L}_3 = \epsilon L_3 = \epsilon \,m \to 0$ as $\epsilon \to 0$ for a given $m$. The vectors $\widehat{L}_a$ with $a=1,2$, satisfy the following commutation relations
$$ \big[ \, \widehat{L}_1\, , \, \widehat{L}_2\,\big] = 0, \quad  \big[ \,  L_3 \, , \, \widehat{L}_1\,\big] = i\widehat{L}_2  \quad {\rm and} \quad \big[ \,  L_3 \, , \, \widehat{L}_2\,\big] = -i\widehat{L}_1 \ , $$
which is the algebra of the inhomogeneous Euclidean group $E_2$. The contraction parameter is arbitrary and has no obvious physical meaning. In the following, we apply IW contraction by identifying $\epsilon$ as the inverse KK-radius to obtain the CSR form the conformal algebra.
\subsection{The CSR from the IW contraction of the Conformal algebra}
We now apply the IW contraction to the generators of the Conformal group. For simplicity, we choose to work in five dimensions. So the light-cone little group is $SO(3)$. In five dimensions, there are three transverse directions. We compactify the third transverse direction, $x^3$, on to a circle of radius $R$,
$$ x^3 = x^3 + 2\pi R \ , $$
and hence the momentum along the third direction become
$$ p^3 = \frac{n}{R} \ , $$
where $n$ is the mode number, called the KK-mode. The light-cone Hamiltonian in Eq.(\ref{mass-shell})now reads,
$$ P^- ~ = ~ \frac{1}{2p^+} \left( p^a p^a + \frac{n^2}{R^2} \right) \ , \quad a=1,2. $$
The term $M^2_n \equiv n^2/R^2$ is the mass term in four dimensions after compactification. We now evaluate all generators in light-cone form at $x^3=2\pi R$. The Poincar\'e generators become\cite{Khan:2004nj}
\begin{eqnarray*}
  M^{+3} &=& -2\pi R \,p^+ \ ,  \\
  M^{3a} &=& 2\pi R\, p^i - \frac{nx^a}{R} + S^{3a} \ , \\
  M^{-3} &=& \frac{nx^-}{R} - 2\pi R\, p^- - \frac{1}{p^+}\left( p^a S^{3a} \right) \ ,\\
  {\rm and} \quad M^{-a} &=& x^- p^a - \frac{1}{2} \{ x^a, P^- \} + \frac{1}{p^+} \left( \frac{n}{R} S^{3a} - p^b S^{ab} \right) \ .
\end{eqnarray*}
The scaling and special conformal transformation generators reduce to the following forms
\begin{eqnarray*}
  D &=& \frac{1}{2} \left( -\big\{ x^- \, , \, p^+ \big\} +  \big\{ x^a \, , \, p^a \big\} \right) + 2n\pi \ ,  \\
  K^+ &=& - x^2_\perp p^+ - 4\pi^2 R^2\,p^+ \ ,  \\
  K^-  &=& 2x^- D + 4n\pi x^-  - x^2_\perp p^-  - 4\pi^2 R^2\,p^-  - \frac{2}{p^+} \left( x^ap^b S^{ab} + 2\pi R\,p^a S^{3a} - x^a \Big(\frac{n}{R} S^{3a} \Big)\right) \ , \\
 K^a &=& 2x^aD - x^2_\perp p^a -4\pi^2 R^2\,p^a + 2x^b S^{ab} - 4\pi R\, S^{3a}  \ , \\
  {\rm and} \quad K^3 & = & 4\pi RD - \frac{n}{R} x^2_\perp - 4\pi^2 n R + 2 x^a S^{3a} \ ,
\end{eqnarray*}
where $a,b=1,2$, and we used $x^2_\perp = x^ax^a$ and also $p^2_\perp = p^a p^a$. The remaining generators do not change.

We now define the light-cone translation vector,
\begin{equation}\label{t-vector}
  T^a ~ = ~ \left(\frac{n}{R}\right) S^{3a} \ .
\end{equation}
Clearly $T^a$'s and the generator $S^{ab}$ satisfy
\begin{eqnarray*}
  \big[ \, S^{ab} \, , \, T^c \,. \big] &=& i\big( \eta^{ac} T^b - \eta^{bc}T^a \big) \ ,  \\
  \big[ \, T^a \, , \, T^b \, \big] &=& \frac{n^2}{R^2} S^{ab} ~ \equiv ~ M^2_n S^{ab} \ .
\end{eqnarray*}
The commutation relations are the $SO(3)$ algebra in four dimensions with mass $M^2_n$, which is known as the regular massive representation. Wigner called these regular massive representations as the `{\it 1st Class}' faithful representations and denoted these by $D^{(j)}$ to classify all irreducible representations of the Poincar\'e group\cite{Wigner:1939cj}.

To contract the algebra, we identify the inverse KK radius, $1/R$, as the contraction parameter. In the limit, $1/R \to 0$, or equivalently $R\to \infty$, the KK mass term vanish, and hence the representation becomes massless. In the contraction limit, the various generators reduce to the following forms,
\begin{eqnarray*}
  \widehat{M}^{-3}  &\equiv & \lim_{R\to\infty} \frac{M^{-3}}{R} ~ = ~ - \frac{1}{np^+} \, p^a T^a \ , \\
  \widehat{M}^{3a}  &\equiv & \lim_{R\to\infty} \frac{M^{3a}}{R} ~ = ~ 2\pi p^a - \frac{1}{n}  T^a \ , \\
  \widehat{M}^{+3}  &\equiv & \lim_{R\to\infty} \frac{M^{+3}}{R} ~ = ~ - 2\pi p^+ \ , \\
  \widehat{M}^{-a} & \equiv  & \lim_{R\to\infty} M^{-a} ~ = ~  x^- p^a - \frac{1}{2} \{ x^a, P^- \} + \frac{1}{p^+} \left( T^a - p^b S^{ab} \right) \ , \\
  \widehat{K}^+ &\equiv & \lim_{R\to\infty} \frac{K^+}{R^2} ~ = ~ -4\pi^2 p^+ \ ,  \\
  \widehat{K}^- &\equiv & \lim_{R\to\infty} \frac{K^-}{R^2} ~ = ~ -4\pi^2 p^-   - \frac{4\pi}{np^+}( p^aT^a) \ ,  \\
  \widehat{K}^a &\equiv & \lim_{R\to\infty} \frac{K^a}{R^2} ~ = ~ -4\pi^2 p^a  - \frac{4\pi}{n}\,T^a \ ,  \\
  \widehat{K}^3 &\equiv & \lim_{R\to\infty} \frac{K^3}{R^2} ~ = ~ 0 \ .
\end{eqnarray*}
The remaining generators that are completely transverse to the compactified direction, such as $M^{ab} = x^a p^b - x^bp^a + S^{ab}$, do not change.  From the above contraction of the algebra, it is clear that, the special conformal translation vector reduces to the linear combination of four-momentum vector $p^\mu$ and the light-cone translation vector $T^a$, and hence can be absorbed by relabelling or shifting the Poincar\'e generators. The dilatation operator picks up a constant term which can be absorbed easily. Since the representation becomes massless after the contraction limits has been applied, it is expected that the Dilatation operator would not vanish. The little group of the contracted representation is generated by the following algebra,
\begin{eqnarray}
  \big[ \, S^{ab} \, , \, T^c \,. \big] &=& i\big( \eta^{ac} T^b - \eta^{bc}T^a \big) \ ,  \label{t-rotation}\\
  \big[ \, T^a \, , \, T^b \, \big] &=& 0 \ . \label{commuting-t}
\end{eqnarray}
The above relation is the algebra of $E_2$ group, and the Casimir eigenvalue is,
\begin{eqnarray}
T^a T^a & = &  \frac{n^2}{R^2} \left( (S^{12})^2 + (S^{23})^2 + ( S^{31})^2 \right) - \frac{n^2}{R^2} (S^{12})^2  \ , \nonumber \\
& = & \frac{n^2 \,j(j-1)}{R^2} - \frac{n^2 m_z^2}{R^2} \ ,  \nonumber
\end{eqnarray}
where $m_z$ is the eigenvalue of $S^{12}$. We now take the limits,
$$R\to\infty, \ \ j\to \infty \quad \mbox{such that } \ \frac{j}{R} \equiv \Xi $$
remains constant. Hence, we obtain, in the contraction limit,
\begin{equation} T^a T^a ~ = ~ \Xi^2 \ . \label{t-length} \end{equation}
The states are now labelled by the values of $m_z$, known as the helicity, and the length of the light-cone vector $\Xi$, and of course by $p^+$ and $p^a$. This is known as the CSR, because a finite boost on a state generates an infinite tower of equally spaced helicity states.

We also obtain an extra term $p^aT^a$ due to contraction. This gives the projection of the light-cone translation vector along the transverse momentum direction. It is interesting to note that $p^a$ transform as a vector under the orbital rotation $L^{ab}$, and $T^a$ transform as a vector under the internal rotation $S^{ab}$ with opposite results,
$$ \big[ \, L^{ab} \, , \, p^cT^c \, \big] ~ = ~ - \big[ \, S^{ab} \, , \, p^cT^c \, \big]  ~ = ~ -i\big( p^bT^a - p^a T^b \big) \ . $$
Hence, the term $p^aT^a$ commute with $M^{ab} \equiv L^{ab} + S^{ab}$ and this means that the orthogonal part of $T^a$ to the momentum vector $p^a$ is physically relevant.

From the above analysis, the generalization to any higher dimension is trivial. The indices of transverse directions now run as $a,b, \cdots = 1, 2, 3, \cdots , (d-3)$. The short little group that label the CSR states is $SO(d-3)$.
\section{Discussion}\label{disc}
In this article we analyzed the IW-contraction of the conformal algebra in $d$ dimensions, and discussed the conditions to obtain the CSR. Since the conformal group contains only the regular massless representations, we intuitively expected that the contraction of the algebra would yield only regular massless representations. Instead, we obtain the CSR in the double limit as both $R, j \to \infty$ such that $j/R = \Xi$ is fixed. Even though the result looks simple, but it is not obvious, because the IW contraction could be applied by compactifying more than one direction. In that case, the contraction may or may not provide the CSR.

Here we have only considered to contract one direction. But from a higher dimension, there are different routes to contract to lower dimensions, and it is not obvious which representation will occur under contraction. It is still an open question. Of particular interest is the AdS/CFT duality. It would very interesting to see the consequence of group contraction on AdS/CFT duality.

We also did not consider the contraction of the algebra generated by the $SO(2,d)$ or $SO(1,d+1)$ in $d$ dimensions. The generators of these  groups also satisfy the same commutation relations of the conformal algebra. It is yet not known if the contraction would yield any CSR or not.
\section{Acknowledgments}
I thank Pierre Ramond for introducing me into this work, and helping me to understand and insisting me to explore further.

\begin{thebibliography}{Ref}
\bibitem{Brink:2002zx} Lars Brink, Abu M. Khan, Pierre Ramond and Xiong, Xiao-zhen, {\it Continuous spin representations of the Poincar\'e and superPoincar\'e groups},  J. Math. Phys., {\bf 43}, 6279 (2002); arXiv: hep-th/0205145.
\bibitem{Weinberg:1980kq} S. Weinberg and E. Witten, {\it Limits on Massless Particles}, Phys. Lett. B {\bf 96}, 59-62 (1980).
\bibitem{Brunetti:2002nt} Romeo Brunetti, Daniele Guido and Roberto Longo, {\it Modular localization and Wigner particles}, Rev. Math. Phys., {\bf 14}, 759-786 (2002); arXiv: math-ph/0203021.
\bibitem{Mourad:2005rt} J. Mourad, {\it Continuous spin particles from a string theory}, arXiv: hep-th/0504118 (2005).
\bibitem{Metsaev:2017ytk} R. R. Metsaev, R.R., {\it Fermionic continuous spin gauge field in (A)dS space}, Phys. Lett. B, {\bf 773}, 135-141 (2017);  arXiv: hep-th/703.05780.
\bibitem{Metsaev:2019opn} R. R. Metsaev, {\it Light-cone continuous-spin field in AdS space}, Phys. Lett. B, {\bf 793}, 134-140 (2019); arXiv: hep-th/1903.10495.
\bibitem{Metsaev:2016lhs} R. R. Metsaev, {\it Continuous spin gauge field in (A)dS space}, Phys. Lett. B, {\bf 767}, 458-464 (2017); arXiv: hep-th/1610.00657.
\bibitem{Khabarov:2017lth} M. V. Khabarov and Yu. M. Zinoviev, {\it Infinite (continuous) spin fields in the frame-like formalism}, Nucl. Phys. B, {\bf 928}, 182-216 (2018); arXiv: hep-th/1711.08223.
\bibitem{Longo:2015tra} Roberto Longo, Vincenzo Morinelli and Karl-Henning Rehren, {\it Where Infinite Spin Particles Are Localizable}, Commun. Math. Phys., {\bf 345}, 587-614 (2016); arXiv: math-ph/1505.01759.
\bibitem{Schroer:2015rct} Bert Schroer, {\it Wigner's infinite spin representations and inert matter}, Eur. Phys. J. C, {\bf 77}, 362 (2017); arXiv: physics.gen-ph/1601.02477.
\bibitem{Kohler:2015cua} Christian K\"ohler, {\it On the localization properties of quantum fields with zero mass and infinite spin}, U. Vienna (main), 2015.
\bibitem{Rehren:2017xzn} Karl-Henning Rehren, {\it Pauli-Lubanski limit and stress-energy tensor for infinite-spin fields}, JHEP, {\bf 11}, 130 (2017); arXiv: hep-th/1709.04858.
\bibitem{Schuster:2013pta} Philip Schuster and Natalia Toro, {\it A Gauge Field Theory of Continuous-Spin Particles}, JHEP, {\bf 10}, 061 (2013); arXiv: hep-th/1302.3225.
\bibitem{Schuster:2014hca} Philip Schuster and Natalia Toro, {\it Continuous-spin particle field theory with helicity correspondence}, Phys. Rev. D, {\bf 91}, 025023 (2015); arXiv: hep-th/1404.0675.
\bibitem{Najafizadeh:2015uxa} X. Bekaert, M. Najafizadeh and M/ R. Setare, {\it A gauge field theory of fermionic Continuous-Spin Particles}, Phys. Lett. B {\bf 760}, 320-323 (2016); arXiv: hep-th/1506.00973.
\bibitem{Najafizadeh:2019mun} Mojtaba Najafizadeh, {\it Supersymmetric Continuous Spin Gauge Theory}, JHEP, {\bf 03}, 027 (2020); arXiv: hep-th/1912.12310.
\bibitem{Buchbinder:2018yoo} I. L. Buchbinder, V. A. Krykhtin and H. Takata, {\it BRST approach to Lagrangian construction for bosonic continuous spin field}, Phys. Lett. B, {\bf 785}, 315-319 (2018); arXiv: hep-th/1806.01640.
\bibitem{Buchbinder:2019iwi} I. L. Buchbinder, S. Fedoruk. and A. P. Isaev, {\it Twistorial and space-time descriptions of massless infinite spin (super)particles and fields}, Nucl. Phys. B, {\bf 945}, 114660 (2019); arXiv: hep-th/1903.07947.
\bibitem{Gracia-Bondia:2017fai} J. M. Gracia-Bondia, F. Lizzi, J. C. Varilly and P. Vitale, {\it The Kirillov picture for the Wigner particle}, J. Phys. A, {\bf 51}, 255203 (2018); arXiv: hep-th/1711.09608.
\bibitem{Bekaert:2017khg} Xavier Bekaert and Evgeny D. Skvortsov, {\it Elementary particles with continuous spin}, Int. J. Mod. Phys. A, {\bf 32}, 1730019 (2017); arXiv: hep-th/1708.01030.
\bibitem{Angelopoulos:1980wg} E. Angelopoulos, M. Flato, C. Fronsdal and D. Sternheimer, {\it Massless Particles, Conformal Group and De Sitter Universe}, Phys. Rev. D, {\bf 23}, 1278 (1981).
\bibitem{Khan:2004nj} Abu M. Khan and Pierre Ramond, {\it Continuous spin representations from group contraction}, J. Math. Phys., {\bf 46}, 053515 (2005) [Erratum: J.Math.Phys. 46, 079901 (2005)]; arXiv: hep-th/0410107.
\bibitem{Yao:1971xg} T. Yao, {\it Unitary Irreducible Representations Of $SU(2,2)$, Reduction With Respect To An Isopoincare Subgroup}, Editors: A. O. Barut and W. E. Brittin, Lect. Theor. Phys. {\bf 13}, 157-174 (1971).
\bibitem{Inonu:1953sp} E. In\"on\"u Eugene P. Wigner, {\it On the Contraction of groups and their representations}, Proc. Nat. Acad. Sci., {\bf 39}, 510-524 (1953).
\bibitem{1921SPAW.......966K} Theodor Kaluza, {\it Zum Unit{\"a}tsproblem der Physik}, Sitzungsberichte der K{\"o}niglich Preu{\ss}ischen Akademie der Wissenschaften (Berlin), 966-972 (1921).
\bibitem{1926ZPhy...37..895K} Oskar Klein, {\it Quantentheorie und f{\"u}nfdimensionale Relativit{\"a}tstheorie}, Zeitschrift fur Physik, {\bf 37}, 895-906 (1926).
\bibitem{AyalaSanchez:2002wm} Mauricio Ayala-Sanchez and Richard Haase, {\it Group contractions and its consequences upon representations of different spatial symmetry groups} in Summer School 2001 on Geometric and Topological Methods for Quantum Field Theory, 435-449 (2002; arXiv: hep-th/0206037.
\bibitem{RevModPhys.21.392} P. A. M. Dirac, {\it Forms of Relativistic Dynamics}, Rev. Mod. Phys., {\bf 21}, 392-399 (1949).
\bibitem{osti_4502829} H. Bacry and N. P. Chang, {\it Kinematics At Infinite Momentum}, Ann. Phys., {\bf 47} 407-423 (1968).
\bibitem{DiFrancesco:1997nk} P. Di Francesco, P. Mathieu and D. Senechal, D., Conformal Field Theory, Springer-Verlag, New York,
    Graduate Texts in Contemporary Physics (1997).
\bibitem{Wigner:1939cj} Eugene P. Wigner, {\it On Unitary Representations of the Inhomogeneous Lorentz Group}, Annals Math., {\bf 40}, 149-204 (1939).
\end{thebibliography}
\end{document}